\documentclass[aps,pre,letter,twocolumn]{revtex4-2}

\usepackage[utf8]{inputenc} % Кодировка UTF-8
\usepackage[T2A]{fontenc}
\usepackage[english,russian]{babel}

\usepackage{graphicx}% Include figure files
\usepackage{dcolumn}% Align table columns on decimal point
\usepackage{bm}% bold math
\usepackage{hyperref}% add hypertext capabilities
\begin{document}

\preprint{APS/123-QED}

\title{\textbf{Accuracy and Limitations of Machine-Learned Interatomic Potentials for Magnetic Systems: A Case Study on Fe-Cr-C}
}%

%%%%%%%%%%%%%%%%%%%%%%%%
\author{E.O. Khazieva}
 \affiliation{Vatolin Institute of Metallurgy of the Ural Branch of the Russian Academy of Sciences, Amundsen str. 101, Ekaterinburg, 620016, Russia}%Lines break automatically or can be forced with \\
\author{N.M. Chtchelkatchev}%
 \affiliation{%
 Vereshchagin Institute of High Pressure Physics, Russian Academy of Sciences, Kaluzhskoe sh. 14, Moscow (Troitsk), 108840, Russia
}%

\author{R.E. Ryltsev}
 \email{rrylcev@mail.ru}
\affiliation{Vatolin Institute of Metallurgy of the Ural Branch of the Russian Academy of Sciences, Amundsen str. 101, Ekaterinburg, 620016, Russia}%
\affiliation{
 Ural Federal University, Lenin Ave, 51, Ekaterinburg, 620002, Russia
}%
\date{\today}% It is always \today, today,

\begin{abstract}
Machine-learned interatomic potentials (MLIPs) have become the gold standard for atomistic simulations, yet their extension to magnetic materials remains challenging because spin fluctuations must be captured either explicitly or implicitly.  We address this problem for the technologically vital Fe–Cr–C system by constructing two deep machine learning potentials in DeePMD realization: one trained on non-magnetic DFT data (DP-NM) and one on spin-polarised DFT data (DP-M).  Extensive validation against experiments reveals a striking dichotomy.  The dynamic, collective properties, viscosity and melting temperatures are reproduced accurately by DP-NM but are incorrectly estimated by DP-M.  Static, local properties, density, and lattice parameters are captured excellently by DP-M, especially in Fe-rich alloys, whereas DP-NM fails.  This behaviour is explained by general properties of paramagnetic state: at high temperature, local magnetic moments self-average in space and time, so their explicit treatment is unnecessary for transport properties but essential for equilibrium volumes.  Exploiting this insight, we show that a transfer-learning protocol, pre-training on non-magnetic DFT and fine-tuning on a small set of spin-polarised data, reduces the computational cost to develop magnetic MLIPs by more than an order of magnitude.  Developing general-purpose potentials that capture static and dynamic behaviors throughout the whole composition space requires proper accounting for temperature-induced spin fluctuations in DFT calculations and correctly incorporating spin degrees of freedom into classical force fields.
\end{abstract}

\maketitle

%\tableofcontents

\section{\label{sec:intro}Introduction}

Magnetic materials are critical to a wide range of technologies, from structural applications to advanced functionalities in spintronics, magnetic refrigeration, and high-density data storage~\cite{Gutfleisch2011AdvMater}. Systems such as transition metal alloys (e.g., Fe-Co-Ni), rare-earth magnets (e.g., Nd-Fe-B), and steels with magnetic phases (e.g., ferritic/martensitic steels) highlight the pivotal role of spin-dependent interactions in governing mechanical, thermal, and functional properties.  Computational design of magnetic materials requires modeling spin interactions across atomic and electronic scales. These simulations demand resource-intensive spin-polarized DFT calculations, restricting their application to systems of just a few hundred atoms. Consequently, DFT cannot capture physical phenomena present in larger-scale structures~\cite{Garrido2025Nanoscale,Tellez-Mora2024NPJCompMater,Kazantseva2008PRB}.

Machine-learned interatomic potentials (MLIPs) have emerged as a transformative tool in computational materials science, bridging the gap between the accuracy of quantum-mechanical methods and the efficiency of classical simulations. Using data from density functional theory (DFT) and advanced regression techniques, such as neural networks, Gaussian processes, or moment tensor potentials, MLIPs achieve near-DFT fidelity in predicting material properties at a fraction of the computational cost. This capability has revolutionized the modeling of complex phenomena, including structural transformations, thermal transport, and chemical reactivity, enabling large-scale simulations of defect dynamics, phase transitions, and interfacial reactions across various material systems~\cite{Cheng2025PRE,Istas2025PRE,Balyakin2020PRE,Wen2022MatFut,Ceriotti2021JCP,Mishin2021ActaMater,VonLilienfeld2020NatComm,Behler2021EPJ,Deringer2020JPhysEnerg,Deringer2019AdvMater,Mortazavi2021AndMater,Tipeev2023JML,Kondratyuk2023JML}.

Recent advances have extended MLIPs to magnetic systems by explicitly incorporating spin degrees of freedom~\cite{Kostiuchenko2024ChiPhysLett,Eckhof2021NPJCompMater,Yang2024PRB,Yuan2024QuanFront,Kotykhov2023SciRep,Kotykhov2024CompMaterSci,Kotykhov2025PRB}. For instance, neural networks with explicit magnetic moment inputs now enable accurate predictions of magnetic ground state structures~\cite{Eckhof2021NPJCompMater}, simulations of spin-lattice coupling in complex crystal structures~\cite{Yang2024PRB} and spin dynamics in low-dimensional magnetic materials~\cite{Yuan2024QuanFront}. Methods such as the Magnetic Moment Tensor Potential have demonstrated success in modeling collinear magnetic states and their coupling to lattice vibrations~\cite{Kotykhov2023SciRep,Kotykhov2024CompMaterSci,Kotykhov2025PRB}.

Thus, developing accurate and general-purpose MLIPs for magnetic materials requires properly accounting for temperature-induced spin fluctuations in DFT calculations and correctly incorporating spin degrees of freedom into classical force fields. The methods to achieve this goal are still in active development and fraught with methodological challenges. Moreover, they require the use of state-of-the-art and extremely expensive first-principles methods to address non-collinear effects and magnetic excited states~\cite{Kostiuchenko2024ChiPhysLett,Cai2023QuantFront}.

An important question arises: Can we use traditional, well-established DFT methods combined with spin-independent machine learning force fields to study the structural, thermodynamic, and transport properties of magnetic metallic alloys? The answer is critically important because, for many practically relevant magnetic materials, such as steels, it is often necessary to simultaneously capture both low- and high-temperature properties, as well as static and dynamic behaviors across wide ranges of compositions and phases. In such cases, the use of advanced techniques for developing magnetic MLIPs is computationally too demanding for most researchers.

Here we address this question considering Fe-Cr-C ternary as a convenient case system. Indeed, Fe-Cr-C serves as a model magnetic alloy due to its complex interplay of competing interactions between ferromagnetic interactions. The interstitial occupation of carbon further modulates these effects, creating a rich landscape of magnetic and structural phases (e.g., martensite vs. austenite stability). From a practical point of view, this system is foundational for stainless steels (corrosion resistance), nuclear reactor materials (radiation tolerance), and bearing steels (wear resistance).
%Magnetic interactions govern key properties of this materials. For example, radiation-induced segregation due to anti-ferromagnetism of Cr or hardness and ductility influenced by Carbon's disruption of Fe-Fe magnetic order. The Fe-Cr-C system thus emerges as an ideal case study, crystallizing both the promise and limitations of MLIPs for magnetic materials, while addressing needs that span fundamental research and industrial optimization.

Thus, we present a comprehensive evaluation of MLIPs for the Fe-Cr-C system, examining their capacity to reproduce structure and physicochemical properties of the system. Our study specifically investigates: (1) the feasibility of developing a unified deep learning potential capable of describing the entire compositional space of Fe-Cr-C alloys, (2) the accurate prediction of diverse properties spanning crystalline and liquid states, including lattice parameters, melting behavior, and viscosity, that are crucial for industrial applications, and (3) strategies to mitigate the prohibitive computational costs associated with spin-polarized DFT sampling. Leveraging the DeePMD framework enhanced by transfer learning techniques, we address how different training protocols provide different accuracy across multiple thermodynamic states. Our findings not only address specific features for the Fe-Cr-C system but also establish general limitations of standard  algorithms to develop MLIPs for magnetic materials.

\section{Deep learning potentials and training procedures}

\subsection{Description of DP models \label{sec:train}}

The interatomic potentials for the Fe--Cr--C system were developed using the DeePMD framework \cite{Wen2022MatFut,DeePMD2_2023JCP}. A key feature of this approach is the use of two coupled multilayer neural networks to approximate the potential energy surface (PES). The first network maps the local atomic environment into a set of symmetry-preserving descriptors, while the second network transforms these descriptors into a local contribution to the total potential energy. Each network consists of three hidden layers with nonlinear activation functions, enabling accurate representation of complex many-body interactions. A more detailed description of the model architecture and training procedure can be found in Ref.~\cite{Wen2022MatFut}. In the following, we will refer to the DeePMD-based potentials as DP (Deep Potential).

Two distinct versions of the potential were constructed for the Fe--Cr--C system. \mbox{DP-NM} (non-magnetic) was trained on configurations obtained from spin-unpolarized density-functional theory (DFT) calculations, whereas \mbox{DP-M} (magnetic) was derived from collinear spin-polarized DFT data. All \textit{ab initio} reference calculations were performed with the Vienna \textit{ab initio} Simulation Package (VASP) using the projector-augmented wave (PAW) method~\cite{Kresse1999PRB} and the Perdew–Burke–Ernzerhof (PBE) generalized-gradient approximation (GGA) exchange–correlation functional~\cite{Perdew1992PRB1}. A plane-wave cutoff energy of 600 eV and generalized Monkhorst-Pack k-point grids~\cite{Wang2021CompMaterSci} generated using KpLib were employed to ensure converged total energies and forces. Spin-polarized calculations for DP-M were initialized with ferromagnetic ordering, allowing both the magnitude and orientation of local magnetic moments to relax self-consistently during the electronic minimization.

\subsection{Active learning procedure for non-magnetic case}

DP-NM and DP-M potentials potentials were generated through two complementary but sequential protocols. In the first stage, the non-magnetic potential \mbox{DP-NM} was constructed by means of an active-learning (AL) workflow~\cite{Wen2022MatFut,Podryabinkin2017ComputMaterSci} implemented in the \textsc{DPGEN} package~\cite{Zhang2020CompPhysCommun}, following the general strategy described in Ref.~\cite{Khazieva2024JCP}.

To seed the AL process, we created 20 supercells (128 atoms each) with disorfered (liquid) structure whose compositions were sampled uniformly across the entire Fe–Cr–C ternary space. Spin-unpolarized \textit{ab-initio} molecular-dynamics (AIMD) simulations in the isothermal–isobaric (NPT) ensemble were performed for these supercells at pressures of 0, 10 and 50~GPa and at a temperature of 3000~K. From the resulting trajectories, the first 20 were retained to form the initial training set; the remaining 80 trajectories were set aside as an external validation set. During each AL iteration, classical MD simulations driven by the current DP-NM potential were executed for the 20 seed configurations. The simulations spanned pressures from 0 to 10~GPa and employed a heating–cooling schedule between 500 and 3000~K, designed to probe both equilibrium and off-equilibrium states. After labeling newly visited configurations with spin-unpolarized DFT energies, forces and virials, half of the seed structures were replaced by the most informative frames (ranked by the DP model’s estimated extrapolation error). Additionally, 350 new labeled configurations were injected into the dataset at every iteration to maintain chemical and structural diversity.

\begin{figure}[t]
\includegraphics[width=0.99\columnwidth]{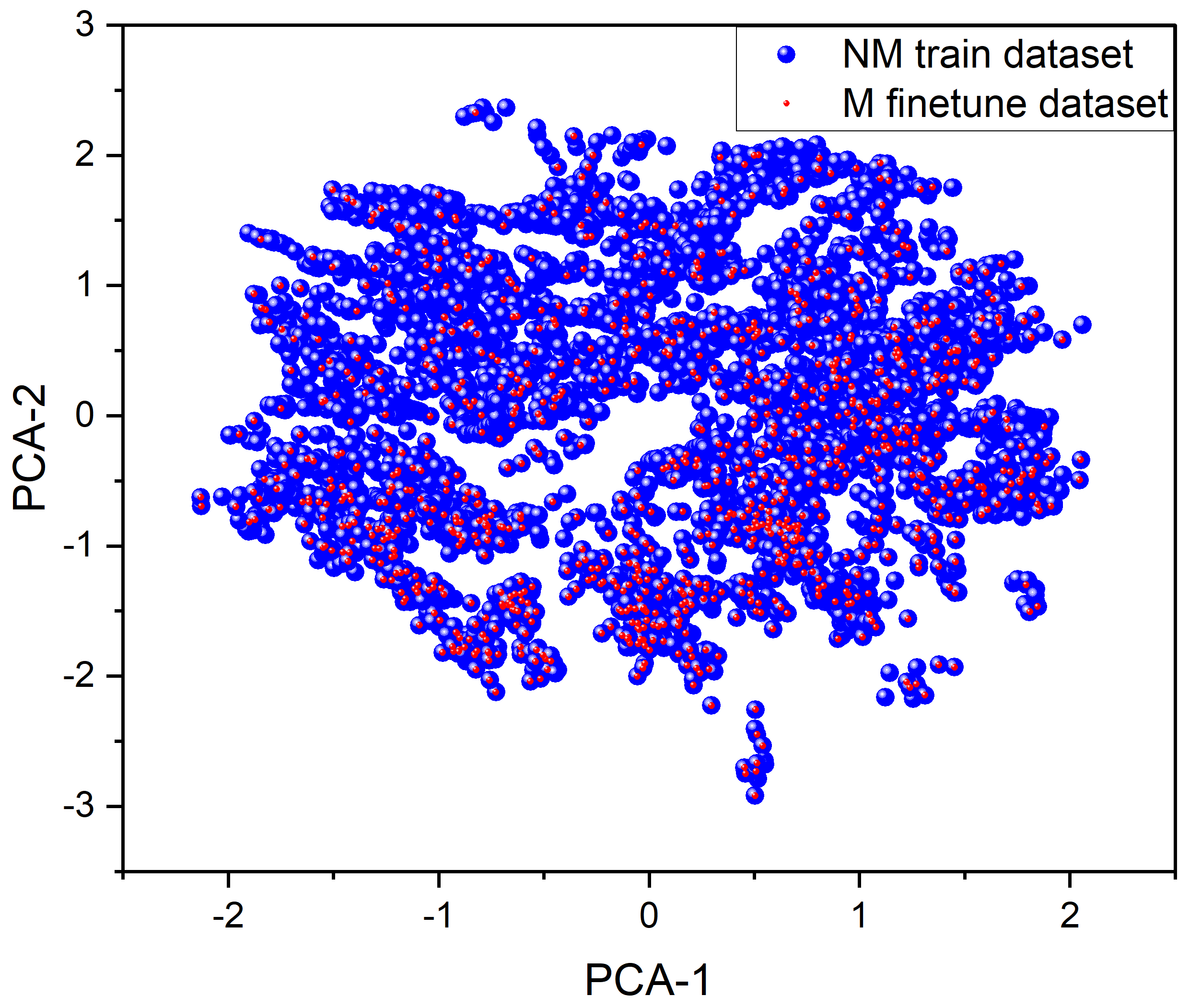}
\caption{Two-dimensional PCA map of structures included in the training datasets. Structures for non-magnetic dataset, constructed by active learning procedure, are represented by blues dots. Structures used for finetuning are presented by red dots.}
 \label{fig:pca}
\end{figure}

After 14 AL iterations, the final database comprised approximately 5000 distinct atomic configurations, each equipped with its DFT total energy, atomic forces and virial tensor. Figure~\ref{fig:pca} presents a two-dimensional embedding of the dataset obtained by principal-component analysis (PCA) of smooth overlap of atomic positions (SOAP) descriptors~\cite{Bartok2013PRB}. The SOAP vectors were computed with the \textsc{ASAP} library~\cite{Cheng2020AccChemRes}. The non-magnetic \mbox{DP-NM} potential was then re-trained from scratch on this curated dataset, yielding a robust and transferable surrogate for subsequent spin-polarized refinements.

\subsection{Transfer learning procedure for magnetic case}

Generating a comparably exhaustive dataset for the magnetic case is far more demanding. Spin-polarised density-functional calculations are roughly twice as expensive per electronic step and, more importantly, exhibit markedly slower self-consistency cycles; the combined effect inflates the required computer time by almost an order of magnitude. To circumvent this bottleneck we adopted a transfer-learning strategy.

A special structure-selection (sparsification sampling algorithm in SOAP space) was applied to the 5000-configuration non-magnetic dataset to extract 150 maximally diverse frames. Each selected frame was subjected to four ionic relaxation steps under spin-polarised conditions, while keeping the cell shape and volume fixed. The resulting ionic positions and final magnetic moments were recorded, yielding 600 labelled structures (4 snapshots $\times$ 150 frames). These configurations are highlighted as red dots in Fig.~\ref{fig:pca}. The pre-trained DP-NM network was used as the starting point for re-optimisation. All weight except ones corresponding to the external layer were frozen, and the network was finetuned on the 600 spin-polarised examples. This procedure produced the magnetic potential \mbox{DP-M}, preserving the broad transferability inherited from \mbox{DP-NM} while endowing it with an explicit, albeit implicit, treatment of collinear magnetism.

\begin{figure}[!]
  \centering
 \includegraphics[width=0.99\columnwidth]{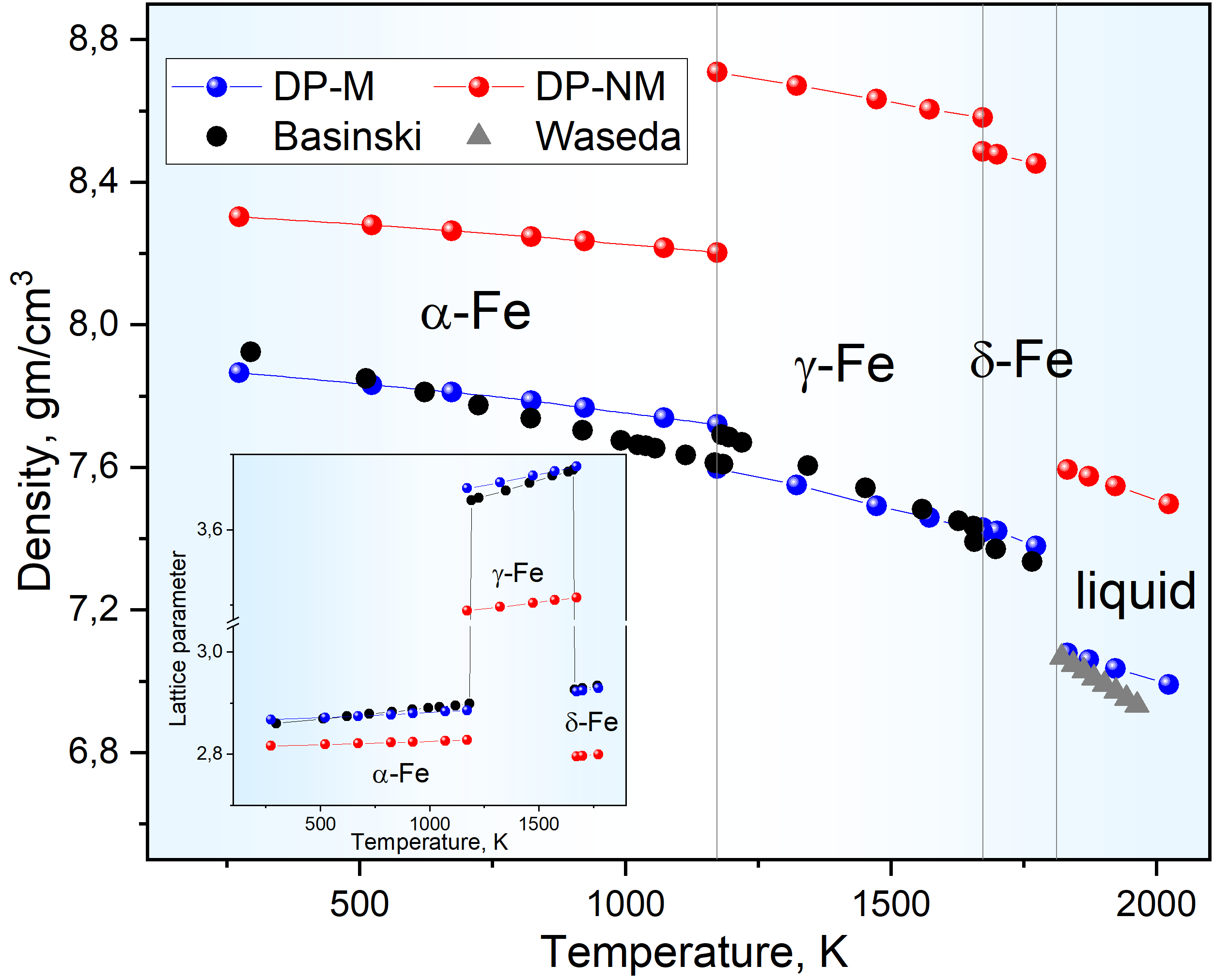}
 \caption{Temperature dependencies of iron density (main frame) and lattice constants (inset). Experimental data are taken from Refs.~\cite{Basinski1955PRS,Waseda1980textbook}.}
 \label{fig:Fe}
\end{figure}

\section{RESULTS}

\begin{figure}
\includegraphics[width=1\columnwidth]{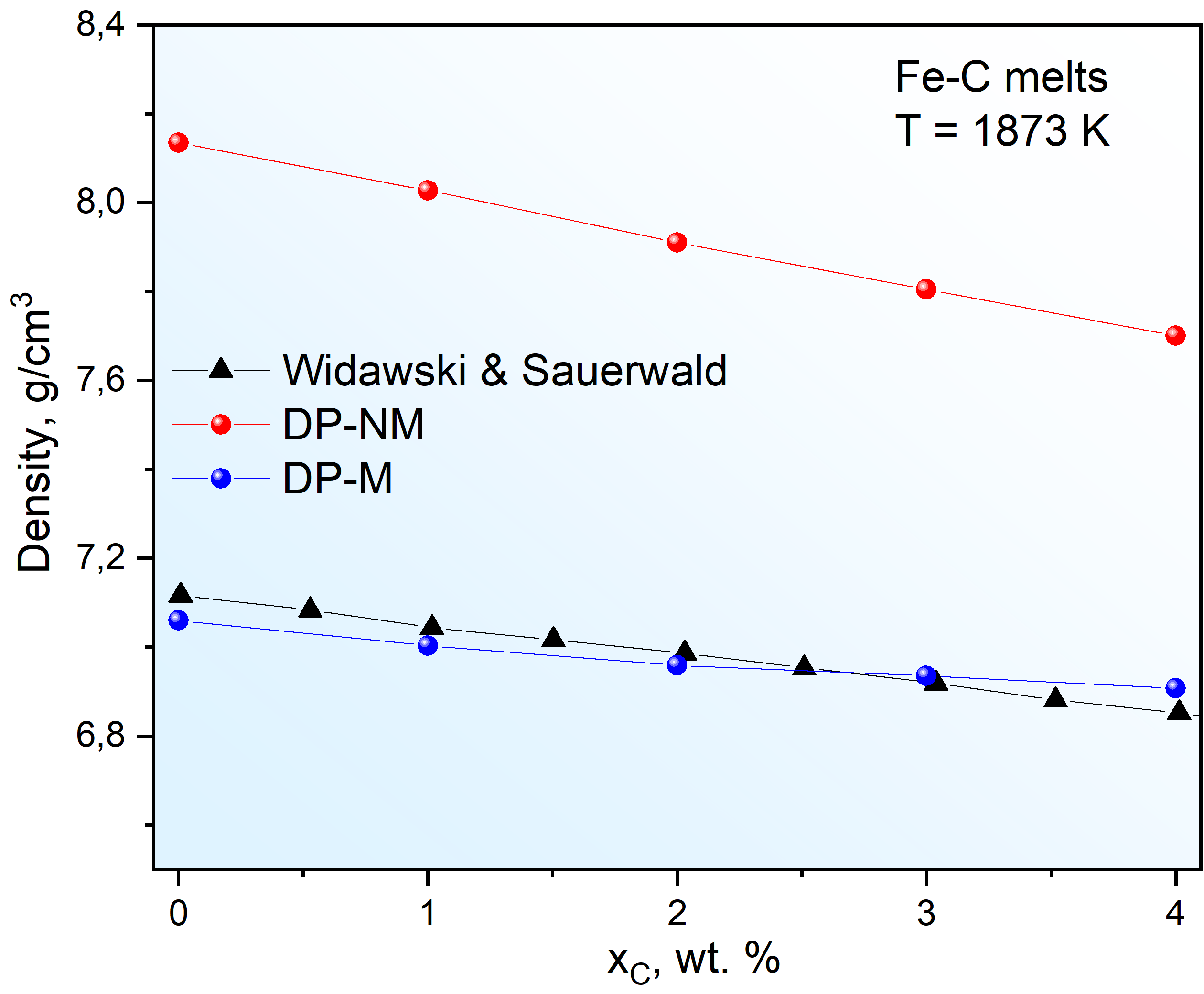}
\caption{Concentration dependence of Fe-C density at 1873 K. Experimental data are taken from Ref.~\cite{Jimbo1993MetallTransB}}
 \label{fig:FeC}
\end{figure}

\begin{figure}
\centering
\includegraphics[width=0.5\textwidth]{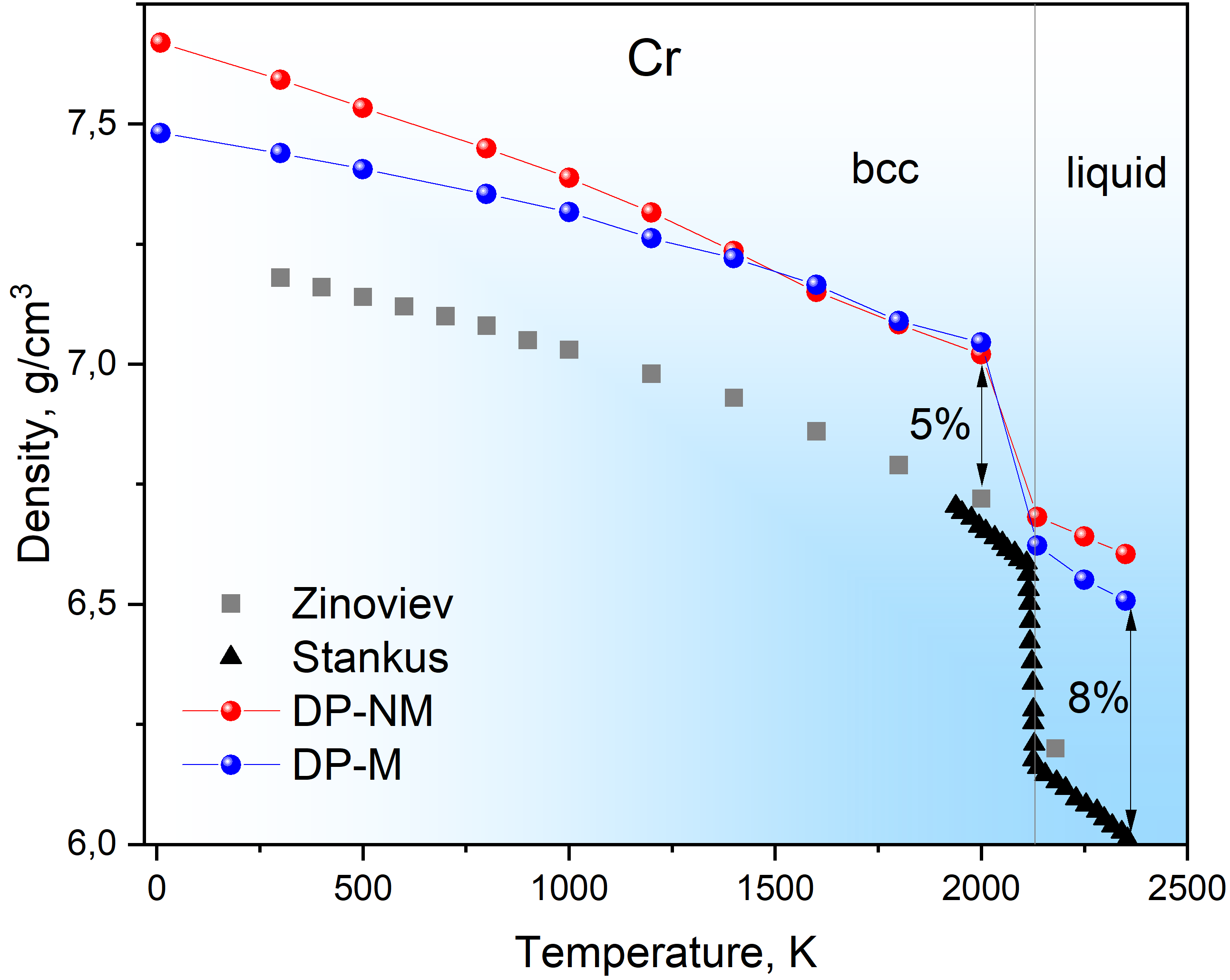}
\caption{Temperature dependencies of chromium density in crystalline and liquid states. Experimental data are taken from Refs.~\cite{Stankus1993TVT,Zinoviev1989}}
 \label{fig:den_Cr}
\end{figure}

In this section, we compare the predictions of the two potentials, DP-M (magnetic) and DP-NM (non-magnetic), for the thermophysical and transport properties of the Fe--Cr--C system. We start with the density and the lattice constants as basic properties that reflect the most general characteristics of interatomic interactions, such as bond strength, equilibrium distances, and thermal expansion behavior, and can be accurately measured in experiment (at least for the solid state). In Fig.~\ref{fig:Fe}, we show both experimental~\cite{Basinski1955PRS,Waseda1980LAS} and simulated temperature dependencies of the density for pure iron; the corresponding dependencies for the lattice constants are presented in the inset. We observe that DP-M provides excellent agreement with experimental data, whereas DP-NM reveals a significant overestimation of the density. A similar situation occurs for the concentration dependence of the density of Fe--C melts at 1873~K (Fig.~\ref{fig:FeC}). The results obtained with DP-NM deviate by approximately 15\% from the experimental data, whereas those from DP-M demonstrate accuracy within about 1\% ~\cite{Jimbo1993MTB}.

Figure~\ref{fig:den_Cr} displays the temperature dependence of the density of pure chromium. Two experimental datasets, covering complementary temperature intervals, are available in the literature~\cite{Stankus1993TVT,Zinoviev1989}; they are mutually consistent within their combined uncertainty. Across the entire temperature range the magnetic potential \mbox{DP-M} reproduces the experimental trend more closely than \mbox{DP-NM}. For the liquid phase both potentials yield a systematic deviation of approximately $8\%$ from the measured values. Upon entering the solid phase the discrepancy of \mbox{DP-NM} grows monotonically with decreasing temperature, whereas that of \mbox{DP-M} remains nearly constant. Even at the lowest temperature examined, the relative error for either potential does not exceed $6\%$, demonstrating that both models maintain quantitative accuracy for the elemental subsystem.

\begin{figure*}[!]
\centering
\includegraphics[width=1\textwidth]{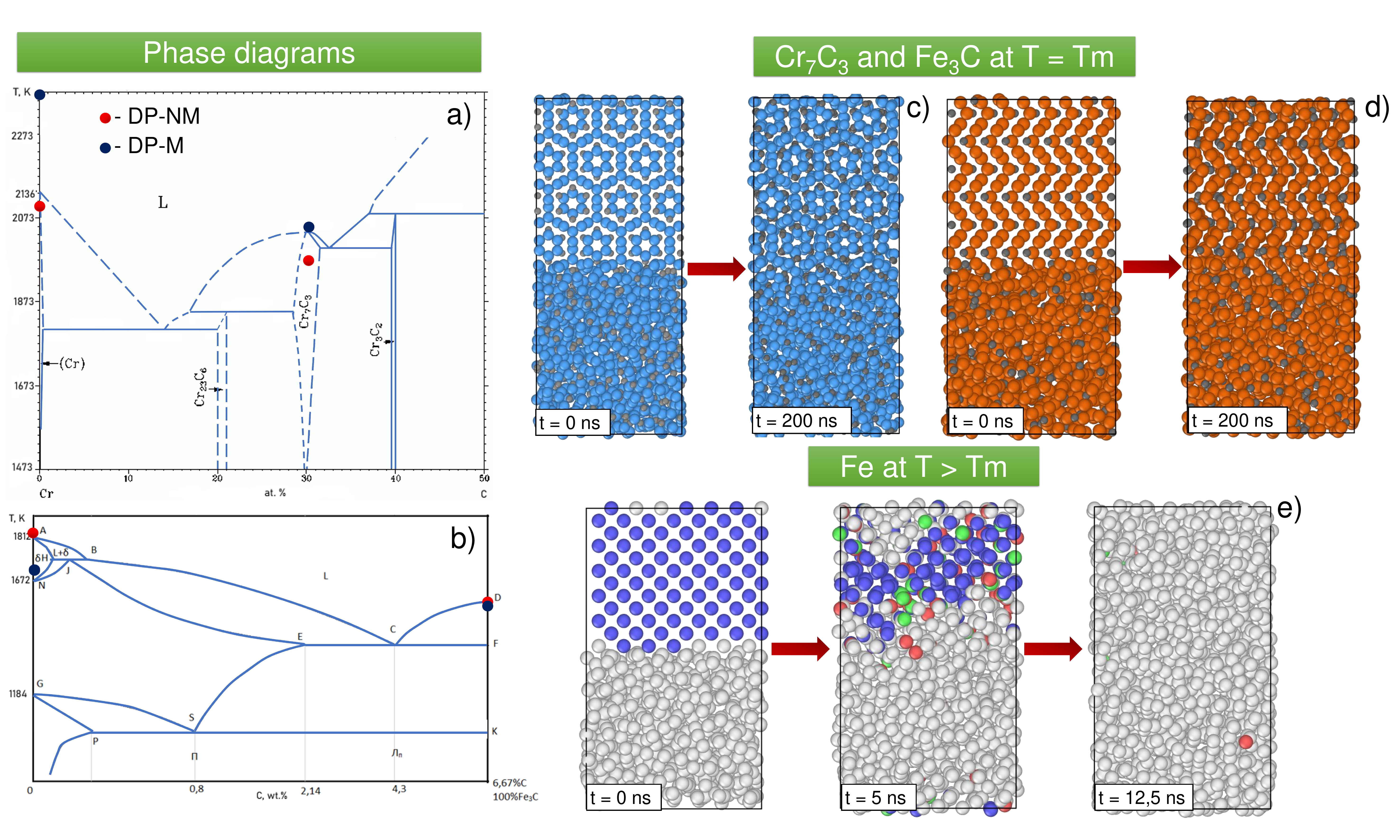}
\caption{(a, b) The sketches of phase diagrams of Cr-C (a) and of Fe-C (b) phase systems. The values of melting temperatures of Fe, Cr, ${\rm Fe_{3}C}$ and ${\rm Cr_{7}C_{3}}$ calculated by DP-NM and DP-M are marked by colored bullets. (c-e) Snapshots corresponding to simulations of movement of the liquid-crystal interface for ${\rm Cr_{7}C_{3}}$ (c),  ${\rm Fe_{3}C}$ (d), and pure Fe (e). The atoms in panels (c,d) are colored via atomic species (blue -- Cr, orange -- Fe, gray -- C) and in panel (e) via local atomic environment as described by polyhedral template matching (blue atoms are bcc-like, green -- fcc-like, and red -- hcp-like, gray -- structureless.) }
 \label{fig:Tm}
\end{figure*}

The next property to be tested is the melting temperature, a more complex quantity because melting is a collective phenomenon—involving the cooperative rearrangement of atoms across phases and the interplay of enthalpic and entropic contributions. We computed the melting temperatures $T_m$ using the two-phase method, in which a solid--liquid interface is simulated at several temperatures and the value at which the two phases coexist in equilibrium is identified.

Four congruently melting phases in the Re--Cr--C system were studied: pure Fe and Cr, and the compounds ${\rm Fe_3C}$ and ${\rm Cr_7C_3}$ (see the Cr--C and Fe--C phase diagrams in Fig.~\ref{fig:Tm}a and b). The melting temperatures obtained with both versions of the potential are indicated by dots; the numerical values are listed in Table~\ref{tab:Tm}. Panels (c) and (d) of Fig.~\ref{fig:Tm} additionally display the two-phase regions for ${\rm Fe_3C}$ and ${\rm Cr_7C_3}$ at their respective melting points. Fig.~\ref{fig:Tm}e shows snapshots of the solid--liquid interface motion in pure Fe at temperatures slightly above $T_m$.

Analysis of the data shows that DP-NM reproduces the experimental melting temperatures for all investigated phases with good accuracy; the deviations remain below 4\%. DP-M yields accurate values for the carbide phases, but it overestimates the melting temperatures of pure Fe and Cr by approximately 10\%.

\renewcommand{\thetable}{\arabic{table}}
\begin{table}
\caption { Melting temperatures of Fe, Cr, ${\rm Fe_{3}C}$ and ${\rm Cr_{7}C_{3}}$ compound experimentally obtained and calculated by simulation of liquid-crystal interface movement using different versions of the DPs.}\label{tab:Tm}
%\fontsize{10pt}{10pt}\selectfont
%\centering
\begin{ruledtabular}
\begin{tabular}{lccccr}
%\hline\hline \\[0.1mm]
 & \multicolumn{3}{c}{Melting temperatures, K}   \\ \cline{2-5} \\[0.1mm]
              &  Fe         & Cr & ${\rm Fe_{3}C}$ & ${\rm Cr_{7}C_{3}}$\\    \hline \\[0.1mm]
experiment    & 1811         & 2136             & 1525     & 2041 \\
DP-NM         & 1820         & 2212             & 1525     & 2000 \\
DP-M          & 1685         & 2368             & 1510     & 2050  \\
\end{tabular}
\end{ruledtabular}
\end{table}

\begin{figure}[t]
\centering
\includegraphics[width=0.45\textwidth]{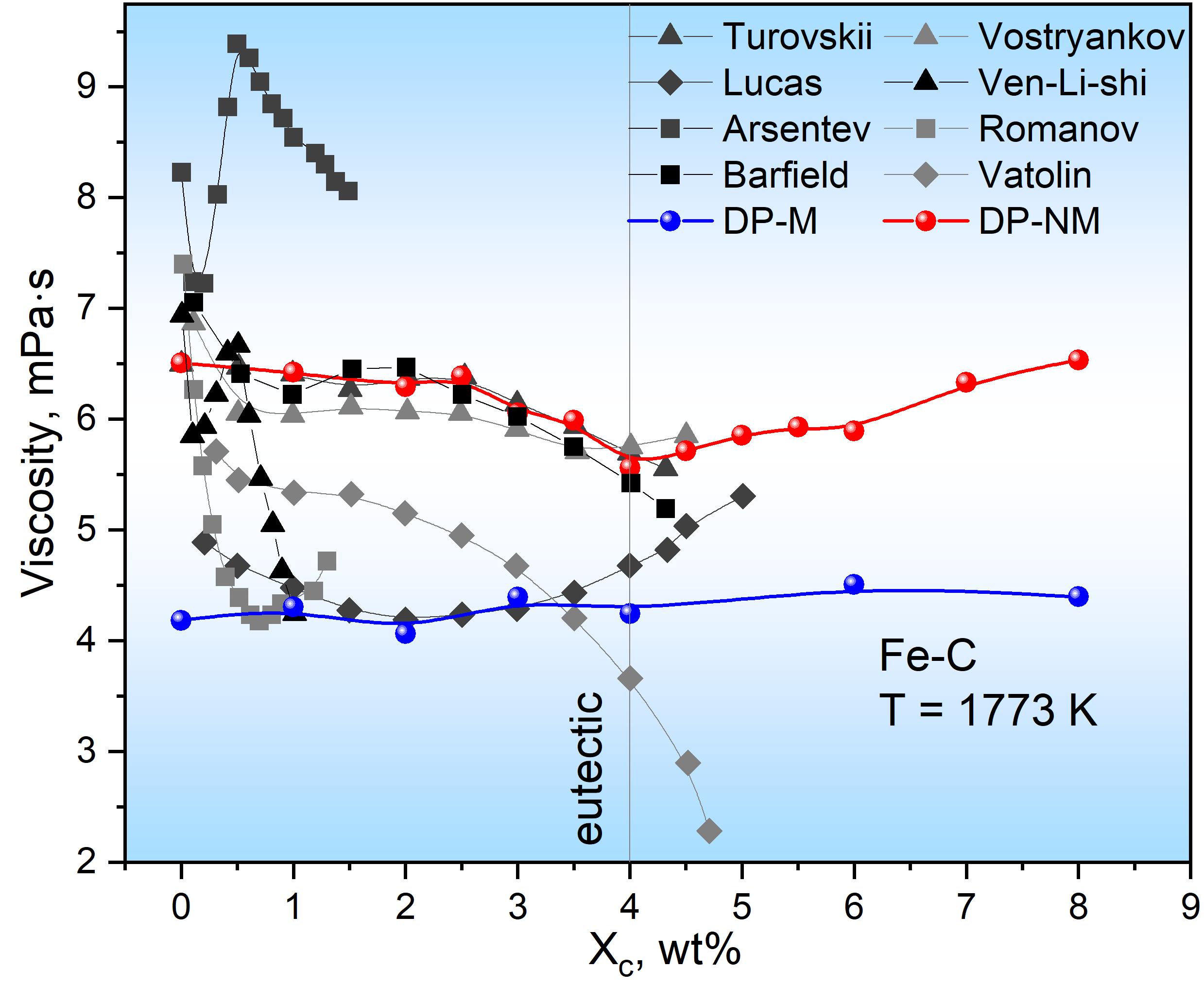}
\caption{Concentration dependence of the dynamic viscosity of Fe-C melts at $T = 1773$ K. Experimental data from Ref.~\cite{Gao2022ChemPhyLett} a represented by grayscale symbols. Lines serve as guides for the eye.
  }
 \label{fig:visFeC}
\end{figure}

\begin{figure}[t]
\centering
\includegraphics[width=0.45\textwidth]{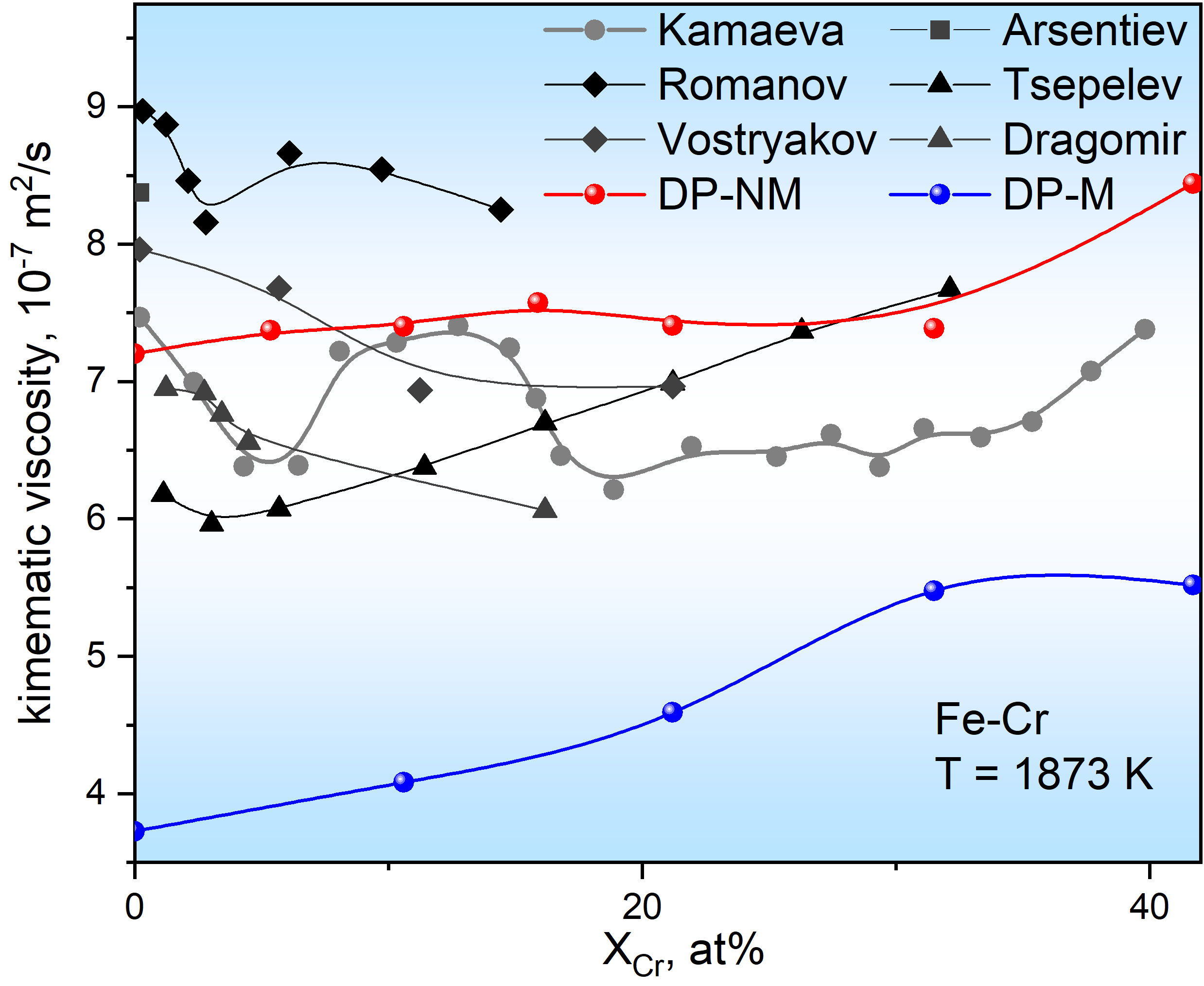}
\caption{Concentration dependence of the kinematic viscosity of Fe-Cr melts at $T = 1873$ K. Experimental data from Ref.~\cite{Kamaeva2012InorgMater} are represented by grayscale symbols. Lines serve as guides for the eye.}
 \label{fig:visFeCr}
\end{figure}

\begin{figure}[t]
\centering
\includegraphics[width=0.45\textwidth]{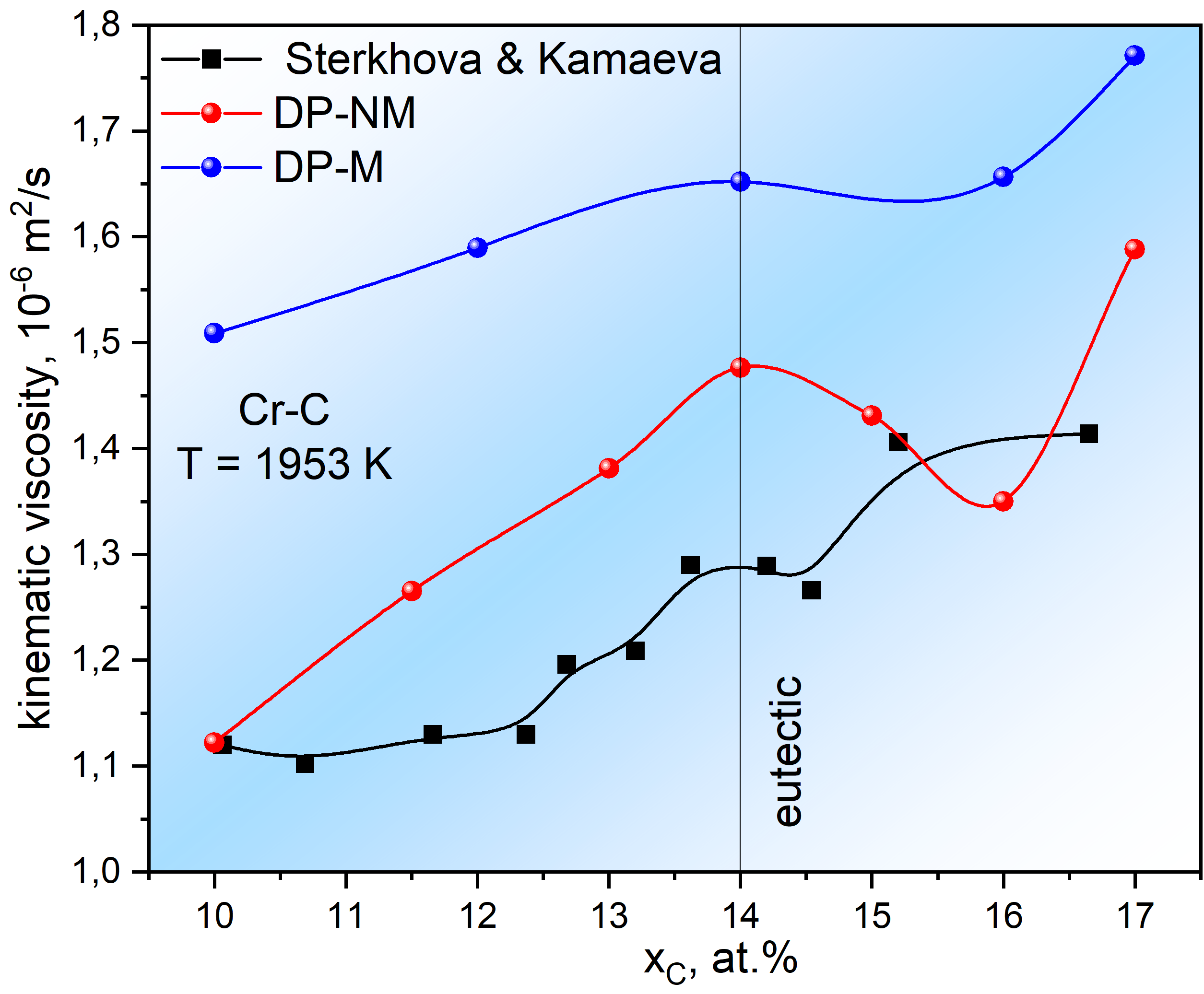}
\caption{Concentration dependence of the kinematic viscosity of Cr-C melts at $T = 1953$ K. Experimental data by Sterkhova and Kamaeva~\cite{Sterkhova2014LNonCrysSol} are represented by black squares. Lines serve as guides for the eye.}
 \label{fig:visCrC}
\end{figure}

Let us proceed to the transport properties of Fe--Cr--C melts, focusing on shear viscosity. It is a key property in many metallurgical processes, yet its measurement for metallic melts demands sophisticated high-temperature experiments. From the computational side, viscosity, which is a collective phenomenon that couples momentum transfer to the evolving melt structure, is extremely sensitive to the precision of the interatomic description and computationally expensive \cite{Khazieva2024JCP,Kondratyuk2023JML}. Hence, simulating viscosity constitutes an ideal stress-test for both the accuracy and the efficiency of an interatomic potential. To address these points we have calculated the concentration dependence of the shear viscosity for the Fe--C, Fe--Cr and Cr--C systems (Fig.~\ref{fig:visFeC}, Fig.~\ref{fig:visFeCr} and Fig.~\ref{fig:visCrC}, respectively).

Let us start with the Fe--C system, which is of central importance in metallurgy as the basis of every steel grade. Figure~\ref{fig:visFeC} presents the calculated concentration dependence of the viscosity at 1773~K together with the available experimental data from Ref.~\cite{Gao2022ChemPhyLett}. Because viscosity measurements for high-temperature metallic melts are challenging, the literature values scatter appreciably; some datasets even differ qualitatively. For example, the data reported by Barfield and Vatolin exhibit an nonphysical trend and are most likely compromised by experimental artefacts. The viscosities obtained with DP-NM are in quantitative agreement with  more recent and reliable experimental studies. The DP-M results reproduce the data of Lucas only up to 3.5~wt\,\% C. Interestingly, DP-M predicts a pronounced minimum at 4~wt\,\% C, i.e.\ exactly at the eutectic composition of the Fe--C phase diagram (Fig.~\ref{fig:Tm}a). A similar correlation between viscosity extrema and the eutectic point has recently been observed in Al--Cu--Ni melts \cite{Kondratyuk2023JML}. In contrast, the DP-NM curve is nearly featureless and does not exhibit any significant extrema.

We now turn to the Fe--Cr system, which is technologically essential because stainless and high-strength low-alloy steels derive their corrosion resistance and mechanical properties precisely from controlled Cr additions. Figure~\ref{fig:visFeCr} shows the calculated viscosity of Fe--Cr melts at 1873~K together with the available experimental data~\cite{Kamaeva2012InorgMater}. The scatter among the latter is even larger than for Fe--C: the reported concentration dependencies differ qualitatively, with some datasets (e.g.\ that of Kamaeva et al.) exhibiting pronounced extrema. Such extrema appear nonphysical in view of the complete high-temperature miscibility of Fe and Cr in both solid and liquid phases, which points to the absence of strong chemical interactions between the two species. Both potentials predict an almost composition-independent viscosity up to $x_{\text{Cr}}\approx 30$~at.\%, followed by a monotonic increase at higher Cr contents. This trend mirrors the smooth variation of the liquidus line in the Fe--Cr phase diagram. Quantitatively, the DP-NM values lie close to the average of the experimental measurements, whereas DP-M underestimates the viscosity by nearly a factor of two across the entire composition range.

%This isotherm shows a sharp increase in viscosity at $x_{\rm Cr}$ = 20 at.\%. In general, both calculated curves increase with increasing carbon concentration. However, this increase in viscosity value occurs in different ways.

Finally, we examine the Cr--C binary system, whose melt properties remain largely unexplored. To our knowledge, only one experimental study by I.Sterkhova and L.Kamaeva~\cite{Sterkhova2014LNonCrysSol} reports the concentration-dependent kinematic viscosity at 1953~K (Fig.~\ref{fig:visCrC}). The DP-NM results are in good quantitative agreement with these data. More important is that both curves demonstrate pronounced features at $x_C=14$ at.\% corresponding to the eutectic composition.  By contrast, DP-M markedly overestimates the viscosity, yet captures the same overall tendency, namely an increase in viscosity with rising carbon content. At the same time, the feature at eutectic composition is much less pronounced.

Across the Fe--C, Fe--Cr and Cr--C systems, the non-magnetic potential (DP-NM) reproduces the measured viscosities far more accurately than the magnetic variant (DP-M). This trend is the reverse of that observed for the density, where DP-M clearly outperforms DP-NM.

\section{DISCUSSION}

The results presented so far demonstrate that the magnetic (DP-M) and non-magnetic (DP-NM) potentials exhibit markedly different accuracies across the various properties and compositions of Fe--Cr--C alloys.
Specifically, DP-M reproduces the density and lattice parameters of iron and iron-rich alloys with excellent precision, yet markedly underestimates/overestimates the melting temperatures of pure Fe/Cr and, above all, yields large errors for the viscosity of all investigated melts. Conversely, DP-NM describes viscosity and melting temperatures accurately, but significantly overestimates the density of iron and iron-based alloys. Notably, both potentials predict the melting points of the carbide phases with comparable accuracy, while neither reproduces the density of pure chromium adequately. These observations raise the question of which physical mechanisms underlie the contrasting performance of the two potentials.

To address this issue we must analyse how different physico-chemical properties are linked to the spatial and temporal distributions of magnetic moments, and how faithfully the two potentials reproduce these distributions. We therefore classify each property according to whether it is governed by (i)~static \emph{versus} dynamic and (ii)~local \emph{versus} collective magnetic characteristics. For example, density is a static, local property: its value is determined by the \emph{instantaneous} local arrangement of magnetic moments around each atom, which modulates the effective interatomic forces and hence the equilibrium volume. Conversely, viscosity is a dynamic, collective property: its magnitude is encoded in the \emph{time-averaged} correlations of the stress tensor, which couples to the magnetic degrees of freedom of the \emph{entire} system (cf.\ the Green–Kubo integral). Similar conclusions can be made for melting temperature.

%Melting temperature occupies an intermediate position: the free-energy difference between solid and liquid depends on both the static magnetic structure that sets the cohesive energy and the long-wavelength magnetic fluctuations that contribute to the vibrational and configurational entropy of the melt.

\begin{figure*}[t]
\centering
\includegraphics[width=0.7\textwidth]{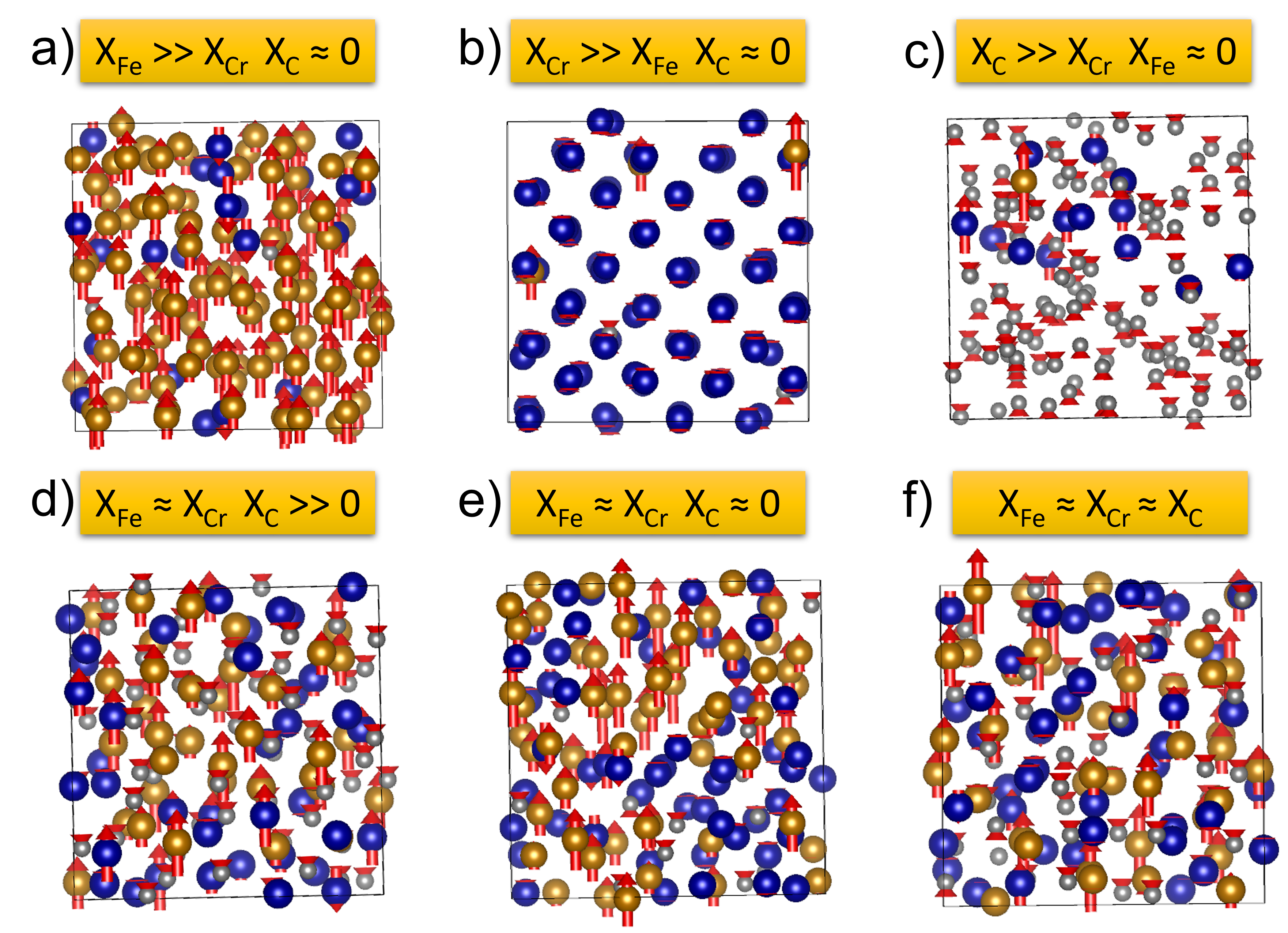}
\caption{Spatial distributions of magnetic moments in  Fe--Cr--C alloys at different compositions.}
 \label{fig:magmoms}
\end{figure*}

The key observation is that, in a paramagnetic system above its magnetic ordering temperature, the local magnetic moments persist with magnitudes close to their low-temperature ferromagnetic values, while their orientations fluctuate rapidly and isotropically. This behavior is captured by the \emph{disordered local moment} (DLM) picture, where the time-averaged moment vanishes due to orientational disorder, but the instantaneous moment remains significant~\cite{Staunton1984JMMM}.

For static properties such as density, the interatomic forces are governed by the instantaneous electronic structure, which responds adiabatically to the slow reorientation of local moments (timescale $\tau_{\rm spin} \sim 10^{-13}$--$10^{-12}$ s). Since the magnitude of the moments remains nearly constant, spin-polarized DFT calculations for collinear configurations, even in a locally ferromagnetic environment, provide an accurate description of the potential-energy surface. Here we also should take into account that the magnetic contribution to interatomic forces depends strongly on the absolute values of magnetic moments whereas its dependence on mutual orientation of the moments is much weaker~\cite{Ruban2007PRB,Ma2015PRB,Rinaldi2024NPJCompMater}. This explains why the magnetic potential (DP-M), trained on such collinear DFT data, performs well for static properties like density. In contrast, dynamic or collective properties (e.g., viscosity, melting temperature) depend on time scales comparable to spin reorientation fluctuations in paramagnetic systems. Non-magnetic DFT fortuitously approximates this by averaging over spin states, allowing DP-NM to provide better agreement with experiment than DP-M, which is based on collinear spin-polarized DFT and so fails to capture disordered spin dynamics.

To illustrate this, Fig.~\ref{fig:magmoms} shows spatial distributions of local magnetic moments in Fe--Cr--C alloys extracted from spin-polarized DFT calculations. For Fe-rich compositions (Fig.~\ref{fig:magmoms}a), the snapshots exhibit nearly ferromagnetic order. The accuracy of DP-M for static properties in this case arises because the high-temperature paramagnetic state can be viewed as an average over uncorrelated, collinear configurations, with forces dominated mostly by absolute values the persistent local moments. For Cr-rich compositions (Fig.~\ref{fig:magmoms}b), the local moments on Cr atoms are negligible, rendering spin-polarized and non-spin-polarized DFT results nearly identical. This is confirmed by Fig.~\ref{fig:den_Cr}, where both DP-NM and DP-M yield similar densities for pure Cr. The residual discrepancies with experiments likely stem from the oversimplification of Cr's actual magnetic ordering (e.g., spin-density waves) in the collinear DFT framework.

The spatial distributions of magnetic moments at intermediate Fe–Cr–C compositions are governed by a subtle competition among ferromagnetic (Fe–Fe), antiferromagnetic (Cr–Cr) and essentially non-magnetic (C-mediated) interactions.  Snapshots c–f in Fig.~\ref{fig:magmoms} reveal that the resulting magnetic landscape is strongly heterogeneous. Standard zero-temperature, collinear DFT calculations capture only a single, static snapshot of this manifold and thus several important effects, shuch as non-collinearity and finite-temperature fluctuations, are therefore missed. Consequently, in general case, potentials trained exclusively on such reference data inherit an incomplete description of the magnetic energetics.

\section{CONCLUSIONS}
In this work we critically examined the widely used protocols that combine either non-magnetic or zero-temperature collinear spin-polarised DFT with machine-learning interatomic potentials (MLIPs) which do not explicitly propagate magnetic degrees of freedom.  Using the technologically central Fe–Cr–C system as a test case, we constructed and benchmarked two DeepMD potentials: one trained on non-magnetic DFT data (DP-NM) and one on spin-polarised data (DP-M). Extensive validation against both high-level \textit{ab initio} results and a broad set of experimental data led to the following principal findings.
\begin{enumerate}
\item \textbf{Dynamic, collective properties}—viscosity and melting temperatures—are described with remarkable accuracy by DP-NM, whereas DP-M systematically underestimates these quantities.  This implies that, \textit{for paramagnetic melts above their magnetic ordering temperature}, the self-averaging of magnetic moments renders an explicit treatment of spin unnecessary for calculating such properties.  Consequently, the development of MLIPs for transport and phase-boundary calculations can be greatly simplified by omitting magnetic degrees of freedom altogether.
\item \textbf{Static, local properties}—density and lattice parameters—are captured excellently by DP-M, especially for Fe-rich (ferromagnetic) compositions.  DP-NM, in contrast, yields large density errors.  Thus, for ferromagnetic alloys, a standard collinear spin-polarised DFT reference at 0~K remains sufficient to train potentials that accurately predict equilibrium volumes and structures up to the liquid state.
\item In Cr-rich regions, where local moments are strongly suppressed or antiferromagnetically compensated, both DP-M and DP-NM give similar (and comparably poor) densities, suggesting that non-collinear and finite-temperature magnetic effects beyond the conventional treatment still need refinement.
\end{enumerate}
Overall, our study delineates a clear practical prescription:
\begin{itemize}
\item Use non-magnetic training sets when the target properties are dynamical or collective;
\item Retain spin-polarised training sets only when high-precision static properties of strongly ferromagnetic phases are required.
\end{itemize}
 This separation not only accelerates potential development (smaller training sets, simpler architectures) but also clarifies the physical origin of the remaining discrepancies.

Additionally, the protocol used to develop magnetic potential offers a decisive advantage in the context of \emph{transfer machine learning}. By pre-training a non-magnetic model on a large, computationally inexpensive non-spin-polarised DFT dataset, and subsequently fine-tuning only a small subset of the parameters with a limited set of spin-polarised calculations, we can transfer the knowledge encoded in the cheap dataset to the magnetic domain. This strategy dramatically reduces the number of expensive spin-polarised DFT runs required to achieve accurate potentials, thereby lowering the overall computational cost by up to an order of magnitude.

Consequently, the combination of our classification scheme (magnetic vs.\ non-magnetic) with transfer-learning workflows opens an efficient route for developing high-fidelity MLIPs for magnetic alloys without the prohibitive overhead of fully spin-resolved training sets. Future work should explore hybrid schemes that properly account for temperature-induced spin fluctuations in density functional theory (DFT) calculations and correctly incorporate spin degrees of freedom into classical force fields.

\section*{ACKNOWLEDGEMENTS}
This work was carried out within the framework of state assignment for IMET UB RAS. The numerical calculations are carried out using computing resources of the federal collective usage center 'Complex for Simulation and Data Processing for Mega-science Facilities' at NRC 'Kurchatov Institute' (ckp.nrcki.ru), supercomputers at Joint Supercomputer Center of Russian Academy of Sciences (www.jscc.ru), 'Uran' supercomputer of IMM UB RAS (parallel.uran.ru).

\bibliographystyle{apsrev4-2}
\bibliography{bib_nnp}% Produces the bibliography via BibTeX.

\end{document}